\documentclass[twocolumn, secnumarabic, amssymb, nobibnotes, prl, superscriptaddress,aps,prl]{revtex4-2}

\usepackage{graphicx} 
\usepackage{amsmath}
\usepackage{float}
\usepackage{color}
\usepackage[english]{babel}
\usepackage{array}
\usepackage[T1]{fontenc}
\usepackage[usenames,dvipsnames]{xcolor}
\usepackage{setspace}
\usepackage{hyperref}
\usepackage{bm}
\usepackage{lipsum} 
\usepackage{tikz}
\usepackage{stmaryrd}

\usepackage{blkarray}
\usepackage{mathtools}
\usepackage{multirow}
\makeatletter
\renewcommand*{\fnum@figure}{{\normalfont \small{FIG.}~\thefigure}}
\makeatother

\DeclareMathOperator{\sgn}{sgn}

\begin{document}

\title{Aging of amorphous materials under cyclic strain}

\author{Dor Shohat}
\altaffiliation{These authors contributed equally to this work}

\affiliation{Department of Condensed Matter, School of Physics and Astronomy, Tel Aviv University, Tel Aviv 69978, Israel}

\author{Paul Baconnier}
\altaffiliation{These authors contributed equally to this work}

\affiliation{AMOLF, 1098 XG Amsterdam, The Netherlands}

\author{Itamar Procaccia}
\affiliation{Sino-Europe Complexity Science Center, School of Mathematics, North University of China, Shanxi, Taiyuan 030051, China}
\affiliation{Department of Chemical Physics, The Weizmann Institute of Science, Rehovot 76100, Israel}

\author{Martin van Hecke}
\affiliation{AMOLF, 1098 XG Amsterdam, The Netherlands}
\affiliation{Huygens-Kamerlingh Onnes Laboratory, Leiden University, 2300 RA Leiden, The Netherlands}

\author{Yoav Lahini}
\affiliation{Department of Condensed Matter, School of Physics and Astronomy, Tel Aviv University, Tel Aviv 69978, Israel}

\begin{abstract}
Amorphous materials driven away from equilibrium display a diverse repertoire of complex, history-dependent behaviors. One striking feature is a failure to return to equilibrium after an abrupt change in otherwise static external conditions. Instead, amorphous materials often exhibit physical aging: an ever-slowing, nonexponential relaxation that can span a huge range of timescales. Here we examine the aging behavior of three different amorphous materials subjected to slow periodic driving. The results reveal a generic aging phenomenon characterized by a logarithmic decay of dissipation per cycle. This observation is evaluated against several mesoscopic models of amorphous matter that successfully capture aging under static conditions: (i) a collection of noninteracting relaxation processes (ii) a noisy hysteron model with random pairwise interactions, and (iii) a structural model consisting of a random network of bi-stable elastic bonds. We find that only the latter model reproduces all experimental findings and relate its success to its persistent, slow exploration of a complex energy landscape with clear signatures of replica symmetry breaking. Thus, cyclic driving emerges as a simple yet powerful protocol to characterize amorphous materials, probe their complex energy landscapes, and distinguish between different models.

\end{abstract}

\maketitle

\section{Introduction}

When driven away from equilibrium, amorphous materials exhibit a range of complex, history-dependent responses, such as critical intermittent dynamics \cite{sethna2001crackling}, anomalously slow relaxations, physical aging \cite{arceri2021statistical}, and a variety of memory effects \cite{keim2019memory}. This diversity of different behaviors, exhibiting complexity over a wide range of spatial and temporal scales, seems to defy a common description.

The out-of-equilibrium response of amorphous materials to external driving is traditionally studied using
two distinct driving protocols \cite{tapias2024bringing}. 
The first protocol explores the long-time response to an abrupt stepwise perturbation. Amorphous materials as diverse as glasses and polymers \cite{arceri2021statistical,kovacs1963glass}, granular packings \cite{knight1995density}, interfaces \cite{ben2010slip,kaz2012physical}, and crumpled materials \cite{matan2002crumpling,lahini2017nonmonotonic} then fail to regain equilibrium and instead exhibit an ever-slowing relaxation process known as 'physical aging', in which the system's properties depend on the time elapsed since the perturbation. This phenomenon is usually described as a sluggish, noise-driven exploration of a complex or hierarchical energy landscape \cite{bouchaud1992weak,matan2002crumpling,robe2016record,shohat2023logarithmic,korchinski2024microscopic}.

The second protocol considers the intermediate time response to 
cyclic driving, which has been used to reveal and investigate self-organization \cite{corte2008random,regev2013onset}, memory effects \cite{paulsen2024mechanical, paulsen2014multiple, fiocco2014encoding, mungan2019networks, shohat2023dissipation}, emergent computational capabilities \cite{bense2021complex,liu2024controlled, kwakernaak2023counting}, and yielding \cite{das2020unified, bhaumik2021role}. As cyclic driving appears to eventually result in a periodic response, these phenomena are usually modeled using athermal quasi-static descriptions \cite{fiocco2013oscillatory}. 

The long term aging behavior during cyclic driving, however, remains under-explored, despite its importance to material properties such as stability \cite{zhao2022ultrastable}, yielding, or failure \cite{parley2022mean,maity2024fatigue}. 

Recent works suggest that in the presence of noise, cyclic driving induces a slow evolution of amorphous systems which converge to a limit cycle in the absence of noise \cite{das2022annealing, priezjev, jana2023relaxation, majumdar2023memory}. Experiments and theory by Bandi \textit{et al.} on cyclic compression of granular materials discovered a power law decay of the energy dissipated per driving cycle \cite{bandi2018training}. An interesting question is thus how generic this behavior is, and what new light can it shed on the modeling of amorphous materials.

Here, we describe the results of applying slow cyclic strain to three very different amorphous materials that exhibit aging under fixed strain: crumpled thin sheets, amorphous bundles of metallic fibers, and shape-memory alloys. We have one main experimental finding and one main theoretical realization. Experimentally, we establish the commonality of response under cyclic drive: these vastly different materials exhibit a remarkably similar aging behavior that has not been previously identified: aging under cyclic drive is characterized by a logarithmic decay in dissipation per cycle. This result, we show, is consistent with the general theoretical framework of Ref. \cite{bandi2018training}.

The theoretical realization is that this behavior allows to discern between different mesoscopic models for physical aging. To this end, we confront our new experimental results with the predictions of models that capture aging under static conditions. These models adopt a coarse-grained view of amorphous materials, describing them as collections of effective, mesoscopic degrees of freedom. We consider three models with increasing complexity and degree of interactions: (i) a non-interacting model assuming independent relaxation processes with a wide Distribution of Relaxation Times (DRT) \cite{macdonald1987linear,chamberlin1998experiments,amir2011huge,amir2012relaxations,lahini2017nonmonotonic}; (ii) a discrete model which allows for the introduction of random, pair-wise interactions (Hysteron model) \cite{preisach1935magnetische,van2021profusion}; and (iii) a structural model describing amorphous materials as a disordered network of bi-stable elastic elements \cite{shohat2025emergent,yan2013glass, shohat2022memory, shohat2023logarithmic}. 

We find that only the last model is consistent with all observations, old and new. We show that the decay of the dissipation per cycle is correlated with a decay in the number of bistable elements that switch back and forth during the cycle. We then connect this to the model's slow exploration of its complex energy landscape, where we demonstrate a dynamical signature of replica symmetry breaking \cite{parisi1983order, mezard1984replica, mezard1987spin, charbonneau2014fractal, scalliet2019nature}. This key factor is absent in the other models, and we discuss the ingredients that seem to give rise to it. Thus, the long time response of aging materials to cyclic driving offers a new characterization of amorphous materials and allows to distinguish between models.

\begin{figure}[t]
\centering
\hspace*{-0.2cm}
\begin{tikzpicture}

\node[] at (0.0,0.0) {\includegraphics[width=\columnwidth]{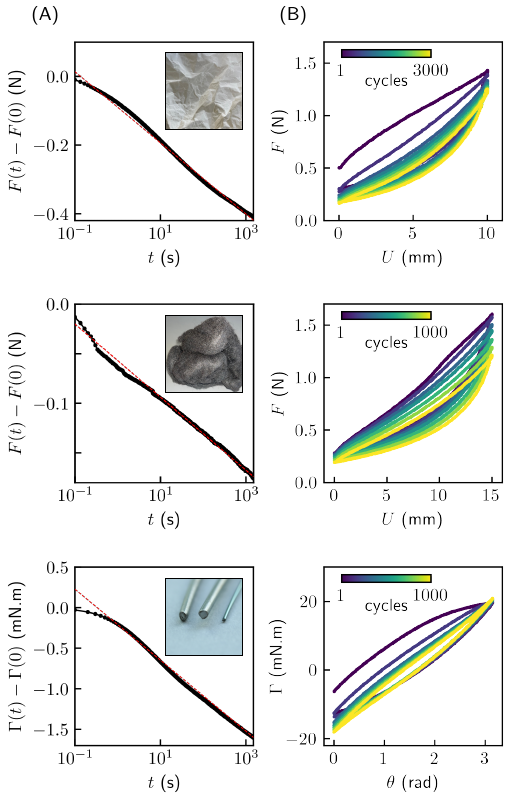}};

\end{tikzpicture}
\vspace*{-0.6cm}
\caption{\small{\textbf{Static and cyclic drive experiments with different amorphous materials.} From top to bottom: crumpled sheet, steel wool, and Nitinol wire in torsion. (A) Force(/torque) relaxation at fixed displacement $U$(/angle $\theta$); the red dashed line represents logarithmic evolution; inset: snapshots of the different experimental systems. (B) Force-displacement curves for repeated cyclic drive, color coded from blue to yellow as time increases; only a few cycles $n$ are shown for clarity: $n = 1,3,10,30,100,300,1000$.}}
\label{fig:poster}
\end{figure}

\begin{figure}[t!]
\centering
\hspace*{-0.45cm}
\begin{tikzpicture}

\node[] at (0.0,0.0) {\includegraphics[width=.51\textwidth]{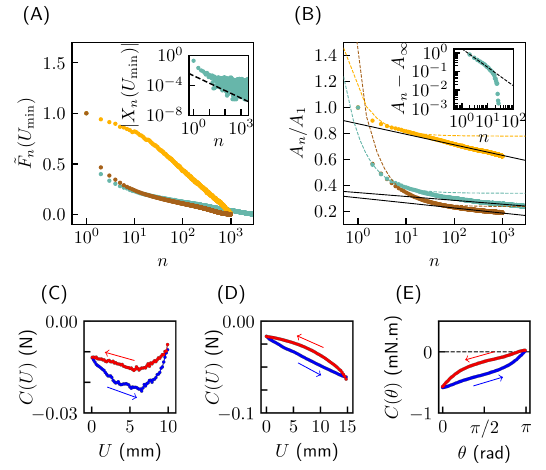}};

\end{tikzpicture}
\vspace*{-0.55cm}
\caption{\small{\textbf{Experimental characterization of cyclic aging.} (A-B) Different colors correspond to different systems, green: crumpled sheet, orange: steel wool, brown: Nitinol wire. (A) Rescaled force $\tilde{F}_n(U) = \left( F_n(U) - F_{\textrm{last}}(U) \right) / \left( F_0(U) - F_{\textrm{last}}(U) \right)$ (torque $\tilde{\Gamma}_n(\theta)$ for the Nitinol wire) at the minimum displacement $U_{\rm min}$ (angle $\theta_{\rm min}$) as a function of the number of drive cycles $n$; (inset) Force increment at the minimum displacement $X_n(U_{\rm min}) = F_{n+1}(U_{\rm min}) - F_{n}(U_{\rm min})$ as a function of the number of drive cycles $n$ for the crumpled sheet. The black dashed line represents $1/n$.
(B) Dissipation per cycle $A_n = \oint F_n(U)dU$ as a function of $n$; the solid black lines represent $A_{0} + \tilde{C}\log n$, where $\tilde{C}$ is obtained from the loop integral of (C-E), and the dashed lines represent $A_{\infty} + a/n$, as obtained from a least square fit; (inset) $A_n - A_{\infty}$ as a function of the number of drive cycles $n$ for the crumpled sheet; the black dashed lines represent $A_{\infty} + a/n$. (C-E) Material creep functions $C(U)$ ($C(\theta)$); blue: ascending, red: descending; C: crumpled sheet; D: steel wool; E: Nitinol wire.}}
\label{fig:overview}
\end{figure}

\section*{Experiments}
\label{sec:experiments} 

We explore three samples of widely different amorphous materials: a thin Mylar sheet crumpled into a ball, an amorphous bundle of metallic fibers (steel wool), and a thin wire of polycrystalline shape-memory alloy (Nitinol in the martensite phase) - see insets in Figs. \ref{fig:poster}A. The crumpled sheet and steel wool are tested using a custom linear compression stage, while the Nitinol wire is tested using a torsional test device (Materials and Methods). Each system is controlled by the imposed displacement $U$ (angle $\theta$ for the Nitinol wire), and we measure the time-dependent force $F(t)$ (torque $\Gamma(t)$) exerted by the system in response.  

To probe the long-time relaxation after a stepwise perturbation, the displacement (angle) is quenched abruptly to a finite value, and the time-dependent force (torque) is measured. All three systems show logarithmic stress relaxation; the force relaxes logarithmically in time for the crumpled sheet (Fig. \ref{fig:poster}A-top and \cite{lahini2017nonmonotonic, shohat2023logarithmic}), for steel wool (Fig. \ref{fig:poster}A-middle), and for the Nitinol wire under torsion (Fig. \ref{fig:poster}A-bottom and Supporting Information). 

Next we consider cyclic driving, where the displacement (angle) oscillates along the interval $[U_{\rm min},U_{\rm max}]$ ($[\theta_{\rm min},\theta_{\rm max}]$), for $n \leq 3000$ cycles (Figs. \ref{fig:poster}B). The three resulting response curves corresponding to the three different materials, $F(U)$ and $\Gamma(\theta)$, show similar dynamics in the form of hysteresis loops with two distinct characteristics. 

First, the force at a given displacement along the $n$-th cycle $F_n(U)$ (or torque at a given angle $\Gamma_n(\theta)$) asymptotically evolves as $F_n(U)\sim C(U)\log n$, where $C(U)$ is a prefactor dependent on the displacement and the drive direction. (Fig. \ref{fig:overview}A). This holds for any strain along the driving cycle, in both the ascending and descending phases (see Supporting Information). The overall response is fully characterized by the function $C(U)$, which we dub the material creep function, which describes the displacement and history dependent logarithmic prefactor (Figs. \ref{fig:overview}C to E). Its shape varies between different amorphous materials, and thus can be used to characterize them. Nevertheless, its hysteretic shape demonstrates a general property of cyclic aging - the force relaxation at each strain is history dependent, and is always faster in the ascending phase.

Second, the area enclosed by the hysteresis loop, indicating the dissipated energy per cycle, decays first as a power law and eventually logarithmically (Fig. \ref{fig:overview}B). While the crossover between these two regimes appears to be material and protocol dependent, the long-time slow relaxation appears to be universal. It is this behavior that we will show to be sensitive to model details, and hence it is central in distinguishing between different models.

\section*{Theoretical argument}
\label{sec:pheno} 

In previous work, Bandi \textit{et al.} \cite{bandi2018training} proposed a general argument for the evolution of the dissipation per cycle $A_n$ for frictional disk packings under cyclic loading (namely force control). The essence of the argument is the existence of an analytic expansion near a fixed point at large $n$, with the leading term in the expansion governing the aging. For force control, this leads to a power-law decay of the dissipation per cycle. Here, we adapt this framework to account for strain control protocols and find that it is consistent with our experimental findings\footnote{We adopt the notation of force and displacement, yet the derivation applies to our torsional measurements as well.}.

We consider the force increment at a given displacement in the $n^{th}$ driving cycle, $X_n(U) = F_{n+1}(U) - F_{n}(U)$. This new variable is history dependent and could be expressed as
$X_{n+1} = g(X_n)$, where the function $g(x)$ is unknown at
this point. $X_n$ must asymptotically vanish, otherwise the force would diverge. Thus $g(x=0)=0$, and near this fixed point a {\em generic} Taylor expansion reads
\begin{equation}
	X_{n+1} = g(X_n) = X_n - K X_n^2 + \cdots
	\label{generic}
\end{equation}

Without an (exceptional) symmetry reason, $K\ne 0$. Therefore, asymptotically, we have $X_n\sim n^{-1}$, and $F_{n}(U)\sim C(U)\log(n)$, where $C(U)$ is the material creep function (see Supporting Information for derivation and leading correction). The dissipation can be obtained by integrating over the response $A_n=\oint F_n(U)dU\sim\log n$, resulting in a logarithmic decay that is consistent with our experimental findings (Fig.~\ref{fig:overview}). One should note that in exceptional cases where $K=0$, higher order terms will lead to different scaling laws.

In contrast to our cyclic strain protocol, the cyclic loading considered in Ref. \cite{bandi2018training} resulted in response curves that did not form closed loops. Instead, their shape allowed an additional approximation for the dissipation, which led to the scaling $A_n\sim1/n$. This shows that aging under cyclic driving is protocol-sensitive.

This reasoning, which is consistent with the observed aging under cyclic driving, is based on a simple and general phenomenological argument, hinting at the generality of our observations. Yet, this description does not provide a microscopic understanding of the mechanisms leading to cyclic aging.

\section*{Searching for a minimal model} \nonumber

To better understand our experimental observations of aging under cyclic drive, we consider three different models that are known to reproduce aging under static drive, and examine their predictions under cyclic driving. These models all adopt a coarse-grained view of amorphous materials, describing them as collections of effective, mesoscopic degrees of freedom. However, the models differ in the choice of these degrees of freedom, and the degree of complexity of their interactions. 

\subsection*{Model (i): Distribution of relaxation times}
\label{sec:DRT} 

Consider first a model of non-interacting degrees of freedom, where slow relaxation under fixed condition is described as the sum response of many exponential relaxations with a wide Distribution of Relaxation Times (DRT). This approach, first used to model the Kolrausch stretched exponential relaxation \cite{macdonald1987linear,chamberlin1998experiments}, was recently shown to account for various aging effects in static and step-wise driving protocols \cite{amir2011huge,amir2012relaxations,lahini2017nonmonotonic}. Below, we show that this description fails to capture the experimentally observed aging during cyclic driving.

Following \cite{lahini2017nonmonotonic}, the DRT model is constructed as follows. We assume a system which is controlled by a single parameter $U$ and which evolves via an ensemble of non-interacting exponential relaxation processes, $V_{\lambda}$, each characterized by a relaxation rate $\lambda$, where the ensemble has a broad distribution of rates $P(\lambda) \propto 1/\lambda$. A key assumption is that for every $U$ there exists an equilibrium state $V^{eq}(U)$, and that all relaxation modes contribute to it equally, i.e. $V^{eq} = \int_{\lambda_{min}}^{\lambda_{max}} P(\lambda)V^{eq}_{\lambda}$, and $V^{eq}_{\lambda} = V^{eq}$. 

Under static conditions, it was shown in previous work that this DRT model exhibits logarithmic relaxation  \cite{amir2011huge,amir2012relaxations}. Namely, from an equilibrium state $V^{eq}_{1}$, following an abrupt change in $U$, its relaxation towards a new equilibrium $V^{eq}_{2}$ is logarithmic in time as long as the observation time is between the physical cutoffs, $1/\lambda_{min}$ and $1/\lambda_{max}$. In subsequent work, the same model was shown to capture the non-monotonic slow response to a two-step driving protocol, leading to slow, Kovacs-like memory responses \cite{lahini2017nonmonotonic}.

\begin{figure}[t!]
\centering
\hspace*{-0.3cm}
\begin{tikzpicture}

\node[] at (0.0,0.0) {\includegraphics[width=\columnwidth]{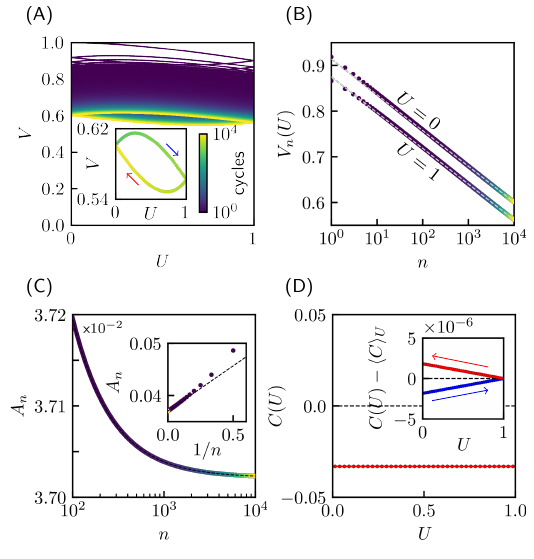}};

\end{tikzpicture}
\vspace*{-0.7cm}
\caption{\small{\textbf{Cyclic aging in simulations of a DRT model.} (A-C) Response to a cyclic drive with fixed $V^{eq}(U) = 1 - U$ and $v = 0.2$. (A) $V(U,t)$; inset: response to the $4$ last cycles.
(B) $V_n(U)$ as a function of $n$ for the minimum ($U = 0$) and maximum ($U = 1$) sweep; the light gray dashed lines represent the large-$n$ logarithmic evolution, as obtained from a least-square method fit.
(C) $A_n$ as a function of $n$; the dashed black line represents $A_{\infty} + \tilde{D}/n$, as obtained from a least square method fit; inset: as a function of $1/n$. (D) Material creep function $C(U)$, evaluated from least-square method fits in $n = \left[9 \times 10^{3}, 10^{4}\right]$; inset: shifted by its average $\langle C \rangle_{U}$; the blue (resp. red) circle markers correspond to $U$ increasing (resp. decreasing);.}}
\label{fig:DRT_simu}
\end{figure}

We now extend the model to consider cyclic driving, by oscillating the control parameter $U$ across a dimensionless interval $U \in \left[ 0, 1 \right]$, at constant speed $v = |dU/dt|$ (Fig. \ref{fig:DRT_simu}A). Each value of $U$ is assigned a different equilibrium state $V^{eq}(U)$, and the relaxation timescales $\lambda$ are distributed in $\left[ 10^0, 10^{6}\right]$ (see Supporting Information). The system is initiated at equilibrium at $U = 0$. Simulations show that the response at a given $U$ along the cycle, $V_n(U)$, evolves logarithmically (Fig. \ref{fig:DRT_simu}B). However, 
the dissipation per cycle, $A_n = \oint V_n(U,t)dU$, decreases as a power-law with the number of drive cycles $n$, such that $A_n \propto 1/n$. (Fig. \ref{fig:DRT_simu}C).

To understand the origin of this power-law decay, we plot the material creep function $C(U)$ as defined above (Fig. \ref{fig:DRT_simu}D). We find that $C(U)$ does not form a closed loop (and neither do the hysteresis loops shown in Fig. \ref{fig:DRT_simu}A). Instead, the gap at $U=0$ dominates aging under cyclic driving. This is very similar to the cyclic loading scenario studied by Bandi \textit{et al.} \cite{bandi2018training}. In the Supporting Information, we also derive the $A_n \propto 1/n$ scaling analytically for the DRT model under sinusoidal driving as long as the observation time is within the physical cutoffs. Altogether, this model is not able to reproduce our experimental observations of a closed loop $C(U)$ and long time logarithmic decay of dissipation under cyclic driving. 

\subsection*{Model (ii): Coupled hysterons}
\label{sec:hysterons} 

The DRT model does not consider interactions between different relaxation modes. These, however, are expected to play a key role in the aging dynamics \cite{shohat2023logarithmic}. To incorporate interactions, we begin by discretizing the model. A discrete version of the DRT model can be constructed by considering $N$ binary and independent relaxation processes, each characterized by a single activation barrier. The distribution of these barriers is taken to be wide, so that at finite temperature, their Arrhenius activation rates are distributed as $P(\lambda)\sim1/\lambda$ \cite{amir2012relaxations}. Such model, still without interactions, was shown to reproduce logarithmic aging in static conditions, while partially capturing the temporal statistics of the intermittencies detected in experiments \cite{lahini2023crackling}.

To consider cyclic driving, we allow the discrete elements to hysterically transition back and forth between their two stable states \cite{shohat2022memory,keim2020global}. In the simplest case, such elements can be modeled as 'hysterons' \cite{preisach1935magnetische,van2021profusion,bense2021complex}. In the following, we consider a collection of noisy hysterons (i.e., at a finite temperature), whose individual parameters are chosen to produce stochastic transition rates that are distributed as discussed above. In addition, we consider the effect of introducing pairwise interactions between the hysterons.

\begin{figure}[t!]
\centering
\hspace*{-0.25cm}
\begin{tikzpicture}

\node[] at (0.0,0.0) {\includegraphics[width=.5\textwidth]{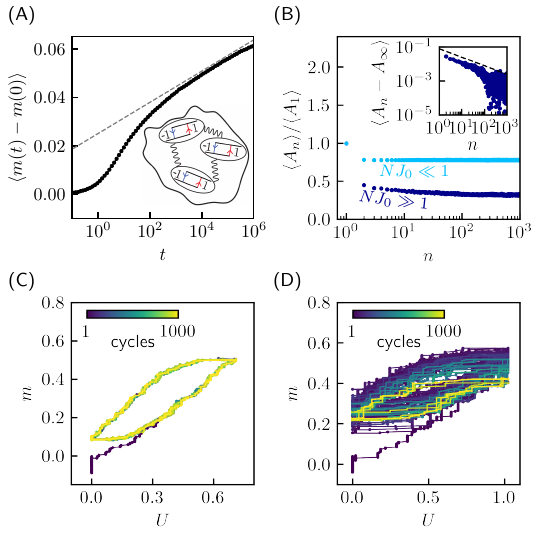}};

\end{tikzpicture}
\vspace*{-0.75cm}
\caption{\small{\textbf{Interacting hysterons at finite temperature  ($T = 10^{-4}$)}. (A) Logarithmic aging under static driving conditions of the  
ensemble averaged magnetization $\langle m(t) - m(0) \rangle$ as a function of time $t$  ($100$ samples, $J_0 = 5. 10^{-4}$, $N=2048$, $U_0 = 0.3$). Inset: schematic of interacting hysterons.
(B-D) Cyclic driving conditions
($N = 512$ and $v = 1.0$). We compare weak interactions ($J_0 = 0.01$, panel B/C) to moderate interactions ($J_0 = 0.077$ , panel B/D). (B) Ensemble averaged, normalized dissipation $\langle A_n \rangle/\langle A_1 \rangle$ as a function of the number of drive cycles $n$ ($200$ realizations). Inset: $\langle A_n - A_{\infty} \rangle$ as a function of $n$, where $A_{\infty}$ is obtained by averaging $\langle A_n \rangle$ for $n \in \left[ 900, 1000 \right]$; the black dashed line represents the slope $1/2$. (C/D) Magnetization-drive curve $m(U)$ for a single realization, for weak (C) and moderate (D) interaction strength.}}
\label{fig:coupled_hysterons}
\end{figure}

We investigate the collective dynamics under static and cyclic drive of ensembles of $N$  hysterons under an external field $U$ (Fig. \ref{fig:coupled_hysterons}A, inset). Each hysteron $i$ is characterized by a state $s_i=\pm1$ and two flipping thresholds $U_i^{-}<U_{i}^{+}$, corresponding to the transitions $s_i : -1 \rightarrow 1$ and $s_i : 1 \rightarrow -1$ respectively. For $U_i^{-}<U<U_i^{+}$ the state is determined by the last threshold crossed. The system is characterized by the state $\boldsymbol{s} = ( s_1, \dots, s_N )$, or by its magnetization $m(t) = \sum_i s_i(t)/N$. In the presence of noise and pairwise interactions, the flipping thresholds can be written as:
\begin{equation} \refstepcounter{equation} \label{eq:coupled_hysterons}
 U_{i}^{\pm}(\boldsymbol{s}, t) = u_{i,0}^{\pm} - \sum_{j \neq i} c_{ij} s_j(t) + \sqrt{2T} \xi_i(t),
\end{equation}
where $u_{i,0}^{\pm}$ are constants termed the 'bare' switching fields, $c_{ij}$ are the interaction coefficients, $T$ is an effective temperature, and the $\xi_i$ are Gaussian white noises with correlations $\langle \xi_i(t) \xi_j(t') \rangle = \delta_{ij} \delta(t-t')$. We take the bare thresholds in a compact range \cite{keim2021multiperiodic, keim2022mechanical, lindeman2023isolating, baconnier2024proliferation}, and we flatly sample the interaction coefficients $c_{ij}$ from $\left[-J_0,J_0\right]$, where $J_0$ is the typical interaction strength. We focus on reciprocal interactions for which 
the interaction matrix is symmetric, i.e. $c_{ij} = c_{ji}$, with zero elements along the diagonal (Materials and Methods). Starting from a random initial condition at $U = 0$, the system is stabilized at zero temperature by flipping unstable hysterons one by one until a stable state is found \cite{hopfield1982neural,baconnier2024proliferation}. The dynamics is then integrated using an event-driven Gillespie-like algorithm (Materials and Methods).

We first consider static drive conditions, namely an abrupt change of the field from $0$ to $U_0 > 0$ at $t=0$. In the non-interacting case, one can show that the flat threshold distribution for $U > U_0$ approximately gives the $P(\lambda)\sim1/\lambda$ flipping rate distribution, and simulations confirm that the magnetization $m(t)$ grows logarithmically in time, with a slope that increases with temperature (see Supporting Information). Finite interaction strengths $J_0$ do not change qualitatively this picture (Fig. \ref{fig:coupled_hysterons}A), up to the emergence of avalanches and the broadening of the distribution of avalanche size as $J_0$ increases, leading to large jumps of magnetization at the level of a single-realization. Eventually, for very large $J_0$ (at fixed $T$), logarithmic aging slows down and ultimately vanishes. This can be traced to the growing spread of the switching thresholds of the hysterons over a range $\propto J_0$, so that eventually thermal fluctuations are no longer sufficient to overcome these barriers. 

Next, we apply cyclic driving by oscillating the field $U$ between $0$ and $U_{\rm max}$ (Material and Methods). First, we consider the case without noise ($T = 0$). 
After $\tau$ cycles the system reaches a limit cycle with a repeatable sequences of hysteron flips \cite{lindeman2021multiple, keim2021multiperiodic, lindeman2024minimal}. Accordingly, the dissipation per cycle $A_n = \oint m(U)dU$ converges to a fixed value. In the small coupling limit $NJ_0 \ll 1$, $\tau = 1$ while for $NJ_0 \gg 1$, $\langle \tau \rangle$ increases with $J_0$ (see Supporting Information).
Second, we consider finite temperature and
weak interactions ($NJ_0 \ll 1$), and drive the system at a fixed speed $v = |dU/dt|$, ensuring that the drive period remains short compared to the relaxation timescales in the absence of driving (Materials and Methods).
We find that $\langle A_n \rangle$ decreases abruptly between the first and the second driving cycle, after which the system gets trapped in a limit cycle and fluctuates around it (Figs. \ref{fig:coupled_hysterons}B and C).
Third, for finite temperature and moderate interactions, 
thermal fluctuations allow exploration of the space of possible limit cycles, and the response then exhibits abrupt jumps of magnetization due to interaction-induced avalanches (Fig. \ref{fig:coupled_hysterons}B and D).
While the relaxation spans multiple drive cycles,  
the mean dissipation per cycle $\langle A_n \rangle$ decays as a power-law $\langle A_n \rangle\sim1/n^\alpha$, with $\alpha=0.7\pm0.15$ (Fig.~\ref{fig:coupled_hysterons}B,D). 
Examining the evolution of individual realizations reveals that these eventually hop between a small number of limit cycles and cease to evolve, as we will see below. 

Altogether, the thermal coupled-hysterons model captures logarithmic aging in static drive conditions, but fails to reproduce our experimental observation of logarithmic decay of the dissipation under cyclic driving conditions. 

\subsection*{Model (iii): Network of bi-stable springs}
\label{sec:bi-net} 
\begin{figure}[b!]
\centering
\includegraphics[width=0.49\textwidth]{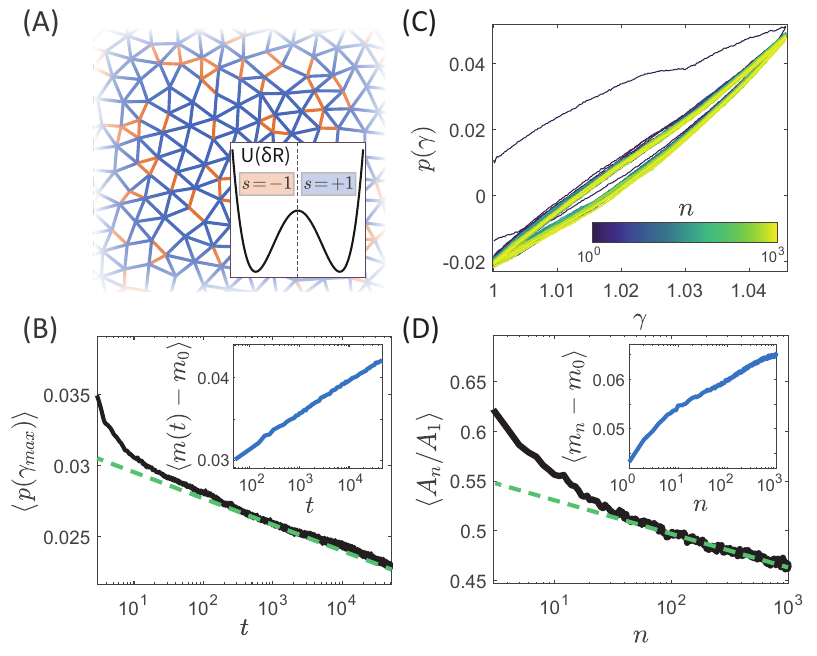}
\vspace*{-0.5cm}
\caption{\small{\textbf{Simulations of bi-stable spring networks.} (A) Illustration of a disordered network with bonds colored by their binary state, red (blue) for $s=-1$ ($s=1$), and the bistable single bond potential as a function of its displacement $\delta R$; (B) Static drive conditions: ensemble averaged pressure $p$ as a function of time $t$ for $N=10^4$ and $T=0.1$ after a rapid increase of strain $\gamma$ from $0$ to $\gamma_{max}=0.045$. The inset shows the change in magnetization over time, which evolves logarithmically as well; (C) Mechanical response $p(\gamma)$ under cyclic driving for a single realization with $T=0.01$; (D) Ensemble averaged, normalized dissipation as a function of the driving cycle $\langle A_n/A_1\rangle$ (same simulation parameters as panel C), well captured by a logarithmic evolution at long times (dashed line). The inset shows the change in magnetization over cycles, evaluated stroboscopically at $\gamma=0$, which evolves logarithmically as well.
}}
\label{fig_binet}
\end{figure}

Finally, we study a structural model composed of a network of bi-stable elastic bonds \cite{shohat2025emergent,yan2013glass} (Fig. \ref{fig_binet}A). This model was recently shown to capture both memory effects under cyclic driving at zero temperature \cite{shohat2022memory}, and logarithmic aging under static loading at finite temperature \cite{shohat2023logarithmic}. In contrast to hysteron models, here the degrees of freedom are continuous. Furthermore, the physical coupling between the bi-stable elements, along with geometric non-linearities, leads to complex interactions that can be non-pairwise \cite{shohat2025geometric}. As we show below, this model successfully captures all of our experimental observations.

Following \cite{shohat2023logarithmic}, we consider disordered networks of $N_b$ bonds, where each bond has a double well potential of the form
\begin{equation}
    U_i=\frac{\alpha_4}{4}\delta R^4-\frac{\alpha_2}{2}\delta R^2~,
    \label{eq:bistable_pot}
\end{equation}
with $\alpha_2$ and $\alpha_4$ network constants, and $\delta R_i$ is the deviation of the $i^{th}$ bond from its randomized rest length $R_i^{(0)}$ (Materials and Methods). Due to the inherent incompatibility between bond lengths, the network carries excess stresses and bonds deviate from the minima of their potentials. This leads to an important difference between this and the previous model: while in the hysteron model the wide distribution of activation threshold (and correspondingly, activations time scales) is prescribed by the thresholds of individual hysterons, here this wide distribution emerges from the self-organized collective state of all bistable springs \cite{shohat2025emergent}.   

We simulate the networks using overdamped Langevin dynamics via LAMMPS \cite{thompson2022lammps} (Materials and Methods). In the simulations, we control the area strain of the network $\gamma(t)=(L^2(t)-L_0^2)/L_0^2$, where $L$ is the system size, while measuring its pressure $p$. A small temperature $T$ serves as external noise \cite{shohat2023logarithmic}. We set $T\ll\Delta U_0$, where $\Delta U_0=\alpha_2^2/(4\alpha_4)$ is the energy barrier for an individual bond. We also define the binarized bond state $s_i=\{-1,1\}$ based on $\sgn(\delta R_i)$, for bonds in their short and long states, respectively. Before applying strain, networks are equilibrated at $T=0$ (Materials and Methods).

Under constant driving, namely when the strain is suddenly increased to $\gamma=0.045$, and $T=0.1$, the pressure exhibits a slow, logarithmic decay \cite{shohat2023logarithmic} (Fig. \ref{fig_binet}B), whose slope increases with temperature (see Supporting Information). Alternatively, one can observe this slow evolution by tracking the binarized magnetization of the system $m=\sum s_i/N_b$. The change in magnetization $m(t)-m(0)$ exhibits a slow logarithmic evolution similarly to the pressure (inset of Fig. \ref{fig_binet}B). 

Next, we subject the networks to bi-axial cyclic strain between $\gamma=0 $ and $\gamma=0.045$. At $T=0$ the mechanical response $p(\gamma)$ reaches a limit cycle after a few oscillations \cite{shohat2022memory}, similarly to the hysteron model. The driving velocity $v=|d\gamma/dt|$ is tuned such that the dissipation $A=\oint p\,d\gamma$ does not strongly depend on $v$ \cite{majumdar2023memory} (see Supporting Information). At finite temperature, however, the dissipation $A_n$ displays a slow, logarithmic evolution over driving cycles $n$, matching our experimental observations (Figs. \ref{fig_binet}C,D). This slow relaxation is also apparent in the evolution of the magnetization $m_n$, evaluated stroboscopically at $\gamma=0$ in each cycle (inset of Fig. \ref{fig_binet}D). In the Supporting Information, we show the material creep function which forms a closed loop, and explore the temperature dependence.

Thus, we can capture the aging behavior under cyclic driving with a structural model of coupled instabilities. The emergence of cyclic aging both in the global mechanical properties (pressure and strain) and in the local state of the bistable units, will now allow us to shed light on its origin.

\begin{figure*}
\centering
\includegraphics[width=0.8\textwidth]{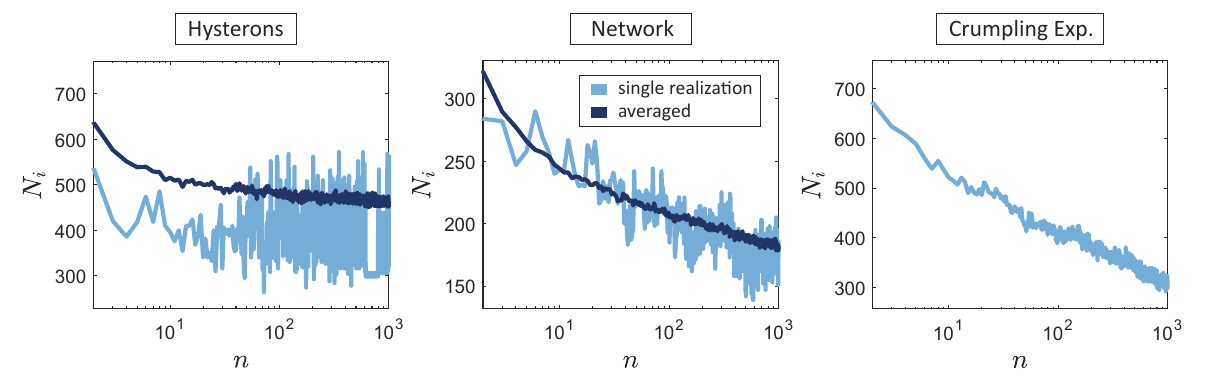}

\caption{\small{\textbf{Structural origin of the decreasing dissipation per cycle.} The latter stems from the decreasing number of instabilities along a full driving cycle $N_i(n)$. For the hysteron model (left, $N = 2048$ hysterons, $J_0 = 0.15$, $T = 10^{-5}$, $100$ realizations), $\langle N_i \rangle$  decreases as a powerlaw with the cycle number $n$. The network model (middle panel, $N_b=5000$ bonds, $50$ realizations) exhibits a logarithmic decrease of $N_i$ with the cycle number $n$, both at the single realization level and on average. In experiments with crumpled sheets (right, single realization) the number of instabilities is measured by counting acoustic emissions during cyclic driving, and exhibits a clear logarithmic decrease.
}}
\label{fig:N_i}
\end{figure*}

\subsection*{Origin of the decrease in dissipation}
\label{sec:N_i}

Equipped with a model that captures our experimental observations, we are now poised to uncover the structural origin underlying the decreasing dissipation over driving cycles \cite{procaccia2025energy}.

We find that the logarithmic decrease of the dissipation per cycle in the hysteron and network models is associated with a similar decrease in the number of elements that flip during a cycle $N_i$ as $n$ increases (Fig. \ref{fig:N_i}-left, Material and Methods). Note, however, that at the level of a single realization, both the dissipation $A_n$ and number of elements involved $N_i$ are not necessarily monotonic with $n$. For coupled hysterons, at long times, we actually observe the system hopping between a few discrete values. On average, consistently with the evolution of the dissipation per cycle, $N_i$ decreases as a power law to an asymptotic value. In stark contrast, in the network model, the number of elements involved $N_i$ decreases logarithmically with $n$, both at the level of a single realization and on average.

Next, we test this prediction in experiment, by measuring the acoustic emissions emitted by the crumpled sheet as it ages under cyclic drive. These acoustic emissions have been shown to the result of the activation of snap-trough instabilities in the sheet, the same instabilities that are modeled here as mechanical hysterons or bi-stable springs \cite{kramer1996universal, Houle1996, shohat2022memory}. Moreover, recent work established that the collective dynamics of these instabilities governs aging under static conditions, without requiring material plasticity \cite{lahini2023crackling,shohat2023logarithmic}.

Using these acoustic measurement, we count the number of instabilities at each driving cycle $N_i$ as a function of the cycle number (Material and Methods). The results are presented in Fig. \ref{fig:N_i}-right. These indicate that, as predicted by the network model, the number of instabilities activated in each cycle decays logarithmically with the cycle number. This suggests the interpretation that similarly to the network model, under cyclic drive the crumpled sheet continues to explore its state space, settling into approximate limit cycles with less and less instabilities, and therefore less and less dissipation.  We note that here, the number of 'clicks' is an underestimation of the true number of instabilities due to the microphone's finite temporal resolution.

\begin{figure*}
\centering
\includegraphics[width=0.8\textwidth]{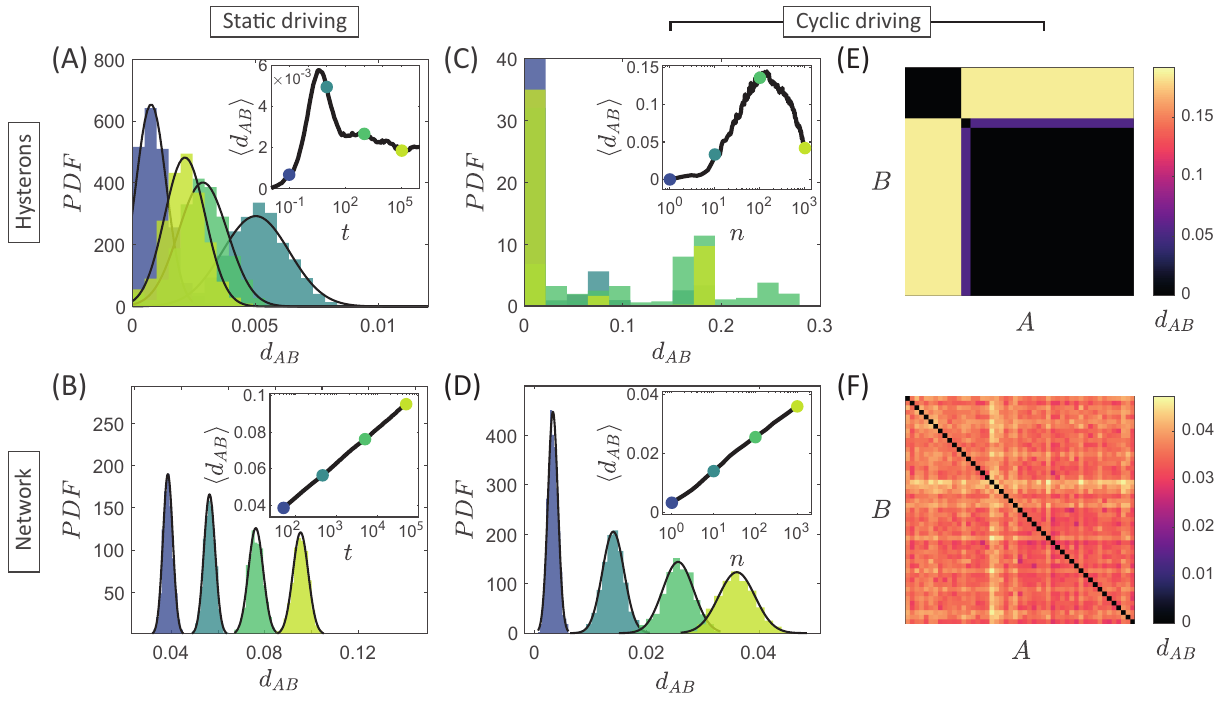}

\caption{\small{\textbf{State space exploration in the different models.} Measurements of coupled hysterons (top), and networks of bi-stable springs (bottom) in the isoconfigurational ensemble. (A-B) Distributions of inter-realization distances $P(d_{AB})$ at different times under static driving, and the mean distance $\langle d_{AB}(t)\rangle$ (inset). Hysterons converge to a fixed distribution while copies of the bi-stable network diverge from each other. (C-D) Distributions of inter-realization distances $P(d_{AB})$ at different cycles under cyclic driving, and the mean distance $\langle d_{AB}(n)\rangle$ (inset). Here $d_{AB}$ is evaluated at the beginning of each cycle. We observe a similar scenario where hysterons converge to small number of limit cycles, while bi-stable networks slowly diverge from each other. (E-F) $d_{AB}$ in a matrix form, comparing all pairs of realizations at $n=10^3$, highlights the difference between the models. 
}}
\label{fig:landscape}
\end{figure*}

\subsection*{Exploring the configuration space during cyclic aging}
\label{sec:isoconfigurational}

The interacting hysteron model and the network model are similar in spirit, introducing interactions between hysteretic two-state degrees of freedom. Nevertheless, while the network model captures all our observations, the hysteron model fails to do so. The network model continues to explore new cycles with reduced dissipation and less participating elements, while the hysteron model eventually remains trapped between a few limit cycles (Fig. \ref{fig:N_i}).
In order to pinpoint the reason for this difference, we measure how the two models explore their state space during the slow evolution.

To this end, we construct an isoconfigurational ensemble of simulations for each model \cite{widmer2004reproducible}. Namely, we run many simulations with the same initial condition, driving protocol, and temperature, but with different realizations of the thermal fluctuations. To quantify how different realizations diverge from each other and explore the space of possible states and limit cycles, we compute the inter-realization distance $d_{AB}=\frac{1}{4N}\sum(\boldsymbol{s}_A-\boldsymbol{s}_B)^2$, where $\boldsymbol{s}_A$ and $\boldsymbol{s}_B$ are the binary states of two copies at a certain time. 

We first compare the evolution of $d_{AB}$ for two models during aging under static drive. For the coupled hysteron model, we find that the mean inter-realization distance $\langle d_{AB}(t)\rangle$ increases, overshoots, and then saturates at a low value (Fig. \ref{fig:landscape}A, inset). In contrast, for the network model $\langle d_{AB}(t)\rangle$ grows logarithmically throughout the evolution (Fig. \ref{fig:landscape}B, inset). In other words, in the network model, copies of the same bistable network diverge from each other under different realizations of the noise. Copies of the same hysteron model follow essentially the same path of states, implying a much simpler organization. This is also visible in the full distribution of inter-copy distances $P(d_{AB})$ (Figs. \ref{fig:landscape}A and B). Although this implies that the mechanism for aging is fundamentally different, both models exhibit similar logarithmic aging under static drive. 


We now consider the two models under cyclic driving. The initial condition for the isoconfigurational ensemble is a limit cycle reached after a transient at $T = 0$ (Materials and Methods). We then switch on temperature, and follow $d_{AB}$ stroboscopically as a function of the number of driving cycles $n$. 
While different realizations of the coupled hysteron model under cyclic driving do not diverge away from each other, and slowly converge to a fixed distribution $P(d_{AB})$ (Fig. \ref{fig:landscape}C), 
we find that copies of the network model diverge logarithmically from each other and $P(d_{AB})$ becomes wider (Fig. \ref{fig:landscape}D). To gain additional insight we represent the set of $d_{AB}$ after many driving cycles in the form of a matrix, where each element represents the distance between two particular copies \cite{scalliet2019nature}. The results reveal that while the hysteron model explores only a small number of limit cycles (Fig. \ref{fig:landscape}E), the network model exhibits a complex organization of the different copies (Fig. \ref{fig:landscape}F), where each realization takes a distinct, separate path in its exploration of the energy landscape.

Formally, the fact that different copies reach different regions of configuration space is a dynamical signature of replica symmetry breaking (RSB) \cite{parisi1983order, mezard1984replica, mezard1987spin, charbonneau2014fractal, scalliet2019nature}. For the hysteron model, the peaked inter-copy distance distribution (Fig. \ref{fig:landscape}C) and the simple organization of the distance matrix (Fig. \ref{fig:landscape}E) imply 1-step RSB, corresponding to an energy landscape with relatively few, well-separated basins. In contrast, the continuous inter-copy distance distribution observed in the network model (Fig. \ref{fig:landscape}D), together with the complex structure of the corresponding distance matrix (Fig. \ref{fig:landscape}F), exhibit characteristics of full RSB, indicative of a substantially more intricate landscape.

These results clearly discriminate between the two models. At long times, the noisy hysteron model becomes confined to a small part of configuration space, and the system converges to a multi-stable steady-state, in which the response alternates between a small set of options. As a result, the transient slow decay of dissipation comes to a halt. In contrast, the bistable spring network continues to explore its configuration space, albeit at a slowing pace, without showing signs of approaching such a steady-state. This exploration allows the system to continuously find cycles with less and less dissipation. 

One important caveat is that both models contain a substantial amount of parameters, and while we cannot rule out that there are parameter ranges or interaction types \cite{baconnier2024proliferation} where the hysteron model would become closer to the network model, at present we have no indications that this would be the case.

\section*{Discussion}
\label{sec:discussion} 

In this work, we have uncovered a new aging phenomena exhibited by different amorphous materials under long-term cyclic driving: a logarithmic decay of the dissipation per cycle. The results were used to discriminate between different models of amorphous matter. In particular, we compared two seemingly similar mesoscopic models: an ensemble of bi-stable degrees of freedom (hysterons) with reciprocal pair-wise interactions, and a structural model of a disordered network of bi-stable elastic elements. While both capture aging under static strain, only the network model captures the full range of experimental results, including the observed aging under cyclic drive. This was shown to result from qualitatively different internal dynamics: only the network model exhibits the dynamical signatures of well-developed RSB. This suggests that cyclic aging arises when the energy landscape is composed of many competing attractors, allowing it to continuously explore phase space without getting trapped. While we believe that static aging originates from a similar mechanism \cite{shohat2023logarithmic}, cyclic aging seems to be more discriminative, possibly due to the numerous intermediate states visited during each drive cycle.

What, then, are the crucial ingredients that enable the network model to more accurately capture the behavior of amorphous matter? We point to two key aspects. First is the continuous nature of the bi-stable degrees of freedom in the network model. As we have recently shown, the frustrated elastic interactions in such networks result in the self-organization of the system into a marginally stable state, in which the barrier for local activations becomes vanishingly small \cite{shohat2025emergent, korchinski2024microscopic}. This persistent property of the network model replenishes the population of low energy barriers and facilitates exploration of the configuration space even at very low temperatures. This property is missing in the coupled hysteron model, where small barriers are depleted over time and the structural evolution comes to a halt. 

The second interesting difference is that the network model includes geometric effects which result in complex non-linear and non-pairwise interactions between the bi-stable elements \cite{shohat2025geometric}. This aspect, absent in the version of the hysteron model studied here, may qualitatively change the complexity of the system's dynamics, and contribute to the slow exploration under cyclic drive. 

Further work is needed in order to establish the necessity of these components for explaining the properties of amorphous media, yet our finding suggests a clear advantage for studying structural models with continuous degrees of freedom, rather than discrete ones.

Aging under cyclic driving can thus serve as a valuable protocol for probing complex behaviors using only global mechanical measurements \cite{shohat2023dissipation}. In particular, the universal logarithmic decay of dissipation per cycle may serve to probe whether the energy landscape of a disordered material is simple or hierarchical \cite{charbonneau2014fractal}. On the other hand, the non-universal Material Creep Function provides a new fingerprint for amorphous materials. These results may have practical implications for experiments in which amorphous matter is cyclically driven, from irreversibility transitions \cite{regev2017irreversibility,reichhardt2023reversible} and memory formation \cite{keim2019memory, keim2021multiperiodic}, to oscillatory yielding and fatigue, which are typically studied at larger amplitudes \cite{bhaumik2021role,das2020unified,maity2024fatigue}. 

Finally, we stress that our approach combines key aspects of glassy physics, notably logarithmic aging, with complex driving protocols as familiar from studies of memory effects. As a wide range of memory effects can be unveiled by more complex driving protocols, i.e. nested driving cycles \cite{sethna1993hysteresis,perkovic1997improved,keim2019memory} or asymmetric driving cycles with time-ordered amplitudes \cite{coppersmith1997self,keim2011generic,paulsen2014multiple,liu2024controlled,lindeman2025generalizing}, we suggest that studying the long time response to such driving protocols in glassy systems at finite temperature may provide additional insight into their underlying self-organization. This may also provide even more stringent tests for models that aim to capture these phenomena.

\section{Acknowledgments}

We thank Oren Raz for fruitful discussions, and Olivier Dauchot for helpful comments. This work was supported by the Israel Science Foundation grant 2117/22.
M.v.H and P.B. acknowledge funding from European Research Council Grant ERC-$101019474$. D.S. acknowledges support from the Clore Israel Foundation. 

\bibliography{bibli.bib}

\appendix
\subsection{Experimental methods} \label{app:experiments}

\subsubsection{Crumpled sheets}

Thin sheets of Mylar, $50$ cm by $50$ cm across and $8\,\mu$m thick, are manually crumpled several times to form a a disordered network of folds and creases. They are loosely formed to the shape of a ball and inserted to the cylindrical custom mechanical tester described in Ref. \cite{shohat2022memory}. The position of the stage $U$ is driven with a precision of $50\,\mu$m, while the normal force $F$ is measured. We note that the forces measured during driving are very small compared to the violent forces applied during the crumpling process. This ensures that the generation of new creases during our experiments is negligible \cite{andrejevic2021model}. 

For static experiments the system is abruptly compressed by $\Delta U=10$ mm at a velocity $v=5\,$mm/s. For cyclic experiments the stage is driven along $U\in[0,10]$ mm at a constant velocity $v=1\,$mm/s. Note that $U=0$ is set by the mechanical tester arbitrarily, and does not represent the free state where $F=0$.

For acoustic measurements, we place the mechanical tester in a sound isolating chamber which reduces external noise by 10 dB \cite{shohat2023logarithmic}. Inside the chamber a microphone records the acoustic emissions, whose waveforms have a typical frequency of 2 KHz. The acoustic signal is processed with a band pass filter, smoothed, and clicks are identified from peaks in the filtered signal. This allows distinguishing between separate events down to a resolution of 0.015 s. 

\subsubsection{Steel wool}

Grade 000\# steel wool, loosely shaped to a ball of diameter $~10\,$cm, is inserted into the same custom mechanical tester. Its edges are not clamped to the tester, but are held firmly between its two plates. It does not lose contact with either plates during driving.

For static experiments the system is compressed by $\Delta U=20$ mm at a velocity $v=5\,$mm/s. For cyclic experiments the stage is driven along $U\in[80,100]$ mm at a constant velocity $v=1\,$mm/s.

\subsubsection{Nitinol wire}

The shape memory wire is made of polycrystalline Nitinol, with activation temperature $T_A \simeq 45^{\circ}$C and with diameter $1$ mm (bought from Smart Wires\copyright). The shape of the Nitinol wire is set straight with a thermal annealing while constraining the wire's shape. This is done by repeating three times $12$ minutes at $500^{\circ}$C following by an abrupt temperature quench in water. The Nitinol wires are then tested in torsion using a torsional INSTRON MT$1$-E$1$. A piece of straight wire of length $L$ is placed into the collet of the testing machine, warmed up above its activation temperature with a heat gun for roughly $30$ seconds, and cooled down for roughly $30$ minutes. For static aging experiments, $L = 22$ mm, and the system is rotated by $360^{\circ}$ at a constant speed $d\theta/dt = 90^{\circ}/s$ before to measure the torque relaxation. For cyclic experiments, $L = 100$ mm, and the system is driven inside $\theta \in \left[ 0, \pi \right]$ at a constant speed $d\theta/dt = 10^{\circ}/s$.

\subsection{Coupled hysterons simulations} \label{app:coupled_hysterons_simulations}

\subsubsection{Sampling}

The midpoints (resp. the span) of the hysterons' hysteresis $\bar{u}_{i,0} = (u_{i,0}^{+} + u_{i,0}^{-})/2$ (resp. $\sigma_i = u_{i,0}^{+} - u_{i,0}^{-}$) are flatly sampled from $\left[-1,1\right]$ (resp. from $\left[0.05,0.5\right]$). The lower bound on the hysterons' spans allows preventing thermal fluctuations from snapping back and forth a given hysteron with a small span.  
Moreover, the interaction coefficients $c_{ij}$ are also flatly sampled from $\left[-J_0,J_0\right]$, where $J_0$ is the typical interaction strength.

\subsubsection{Numerical integration}

For noiseless cyclic drive simulations, we use an event-driven algorithm introduced in \cite{keim2021multiperiodic} to simulate the sequence of hysteron flips as the input $U$ is varied quasistatically. Similarly, for $T > 0$, we integrate the dynamics by moving forward in time instability-by-instability using the approach defined in \cite{korchinski2022dynamic}, further described below. Not only are event-driven methods very efficient computationally-wise, but they also allow to make sure instabilities are systematically triggered one by one.

At a given time $t$, hysteron $i$ has a distance to instability $\Delta U_i(t)$, and flips by thermal activation at rate $\lambda(\Delta U_i) = P(\xi_i > \Delta U_i)$ or immediately at $\Delta U_i = 0$. Working out the inter-event period $\delta t$ and site $i$ is relatively trivial for fixed $U$, as activations happen at a given rate $\lambda(\Delta U_i)$. Since thermal activations are independent processes, with a waiting time probability distribution function $p(t) =
\lambda(\Delta U_i)e^{-\lambda(\Delta U_i)t}$, for a system with $N$ hysterons one could sample $N$ random numbers $\{R_i \in \left[0, 1\right)\}$ and find the next activation time for each hysteron by inverting the cumulative distribution function for the waiting times, so that each hysteron is assigned a time $t_i = -\frac{1}{\lambda(\Delta U_i)} \ln(1 - R_i)$. By finding the hysteron with the smallest $t_i$, one has found the first hysteron to flip, and the appropriate interval $\delta t = t_i$. However, in cyclic drive simulations, the rates $\lambda(\Delta U_i(t))$ are not constant: $\Delta U_i(t) = \Delta U_i(t=0) \pm vt$ (depending on the drive's direction and the state of hysteron $i$). In this case, for each hysteron we sample a random number $R_i \in \left[0, 1\right)$, and solve for $t_i$ as:

\begin{equation} \label{eq:event_driven_stochastic_hysterons}
 R_i = P(t < t_i) = \exp \left[ - \int_0^{t_i} \lambda(\Delta U_i (t)) dt \right].
\end{equation}

To solve Eq. (\ref{eq:event_driven_stochastic_hysterons}) numerically, consider the related function $R(t) = \exp \left[ - \int_0^t \lambda(\Delta U_i (t'))dt' \right]$, for which $R(t_i) = R_i$ is the desired solution. $R(t)$ obeys the following nonlinear ODE:

\begin{equation}
\frac{dR}{dt} = \left[ 1 - R(t) \right] \lambda(\Delta U_i(t)),
\end{equation}

for which the initial value problem $R(t = 0) = 0$ and $R(t_i) = R_i$ (where $R_i$ is still randomly drawn from $\left[0,1\right)$) has a unique solution. We use SCIPY’s solve-ivp routine \cite{virtanen2020scipy} to integrate these equations for all hysterons simultaneously, halting when either $R(t_i) = R_i$ or $\Delta U_i = 0$ for any hysteron.

The system's preparation by stabilizing a random initial condition is performed at zero temperature. In static conditions, the quench by $\Delta U$ is simulated using the noiseless event-driven algorithm, so that the switching fields are considered fixed throughout the quench. However, from $t = 0$, the dynamics is computed using the stochastic event-driven simulation scheme. In cyclic drive conditions, the maximum drive amplitude $U_{\textrm{max}}$ is determined during the first drive cycle by driving the system up to a magnetization $m = 0.5$, and keeping this maximum drive amplitude fixed for the rest of the simulation.

\subsubsection{Isoconfigurational ensemble} The ensemble is prepared by creating one instance of the model following the sampling described above, and by driving cyclically this single instance repeatedly at $T=0$ until reaching a limit cycle. Multiple copies of the system are then stored, and simulated independently in the presence of noise for $n = 1000$ cycles.

\subsubsection{Number of instabilities} Here, we define the number of instabilities over the $n$-th driving cycle $N_i(n)$ as the sum of the numbers of elements which have flipped during the increasing and decreasing phases. The latters are determined by the number of elements which changed state, simply comparing the system's configurations at the lowest and largest values of the drive.

\subsection{Bistable spring network simulations} \label{app:bistable_networks_simulations}

We follow the simulation scheme detailed in Ref. \cite{shohat2023logarithmic}. Over-coordinated network topologies are obtained from jammed packings as detailed in Ref. \cite{shohat2022memory}. We use large networks with $N=2000$ and $N=10000$ nodes ($N_b\approx5000$ and $N_b\approx25000$ bonds respectively). Each bond is assigned with random rest length $R_i^{(0)}\in[8,11]$, and a potential given by Eq. \ref{eq:bistable_pot} for deviations from the rest length $\delta R_i=R_i-R_i^{(0)}$. For all bonds the potential is identical with $\alpha_2=2.5$ and $\alpha_4=1$.

We use the molecular dynamics platform LAMMPS \cite{thompson2022lammps}. Brownian dynamics are simulated with a Langevin thermostat, an integration timestep $dt=0.005$, a damping coefficient $\beta=0.5$, and varying temperatures $T$. We use fixed boundary conditions. Cyclic bi-axial strain is realized by setting a constant velocity $v$ to the top and right boundaries of the network in the directions $\hat{y}$ and $\hat{x}$ respectively. The bottom and left boundaries are pinned in $\hat{y}$ and $\hat{x}$ respectively, and are free along the perpendicular axes. Before applying strain, we prepare each network at $T=0$ and constant strain to reach a local mechanical equilibrium. The magnetization $m=\sum_i s_i/N_b$ at the beginning of cyclic driving, which we denote $m_0$, can be tuned by applying a constant force $F_{prep}$ to the network's boundaries during preparation at $T=0$. This also sets the initial system size $L$.

For static aging simulations we start by an abrupt change in strain from zero to $\gamma_{max}=4.5\%$ at a velocity $v=2$ and $T=0$. We then turn on temperature and allow the networks to evolve.

\end{document}